\documentclass[12pt] {article}
\usepackage{amsmath,latexsym}
\usepackage[ansinew]{inputenc}
\usepackage{amsmath}
\usepackage{amssymb}
\usepackage{graphicx}
\usepackage{color}
\usepackage[colorlinks]{hyperref}
\begin{document}

\title{\textbf{On the role of pressure anisotropy for relativistic stars admitting conformal
motion}}
\author{\small Farook Rahaman$^\heartsuit$\footnote{Corresponding author: farook\_rahaman@yahoo.com},
 Mubasher Jamil\footnote{mjamil@camp.edu.pk}, Mehedi Kalam\footnote{mehedikalam@yahoo.co.in},
 Kaushik Chakraborty$^\heartsuit$ and  Ashis Ghosh$^\heartsuit$\\ \\
%EndAName
$^\heartsuit$\small Dept. of Mathematics, Jadavpur University,
Kolkata-700 032,
 India\\ \\
 $^\dag$\small Center for Advanced Mathematics and Physics,
National University of Sciences and Technology,\\ \small Peshawar
Road, Rawalpindi - 46000, Pakistan
\\ \small and \\
$^\ddag$\small Department of Physics, Netaji Nagar College for
Women, Regent Estate, Kolkata-700092, India
%EndAName
 }\maketitle

 \begin{abstract}
We investigate the spacetime of anisotropic stars admitting
conformal motion. The Einstein field equations are solved using
different ansatz of the surface tension. In this investigation, we
study two cases in details with the anisotropy as: [1]  $p_t = n
p_r$ [2] $p_t - p_r = \frac{1}{8 \pi}( \frac{c_1}{r^2} + c_2)$
where, n, $c_1$ and $c_2$ are arbitrary constants. The solutions
yield expressions of the physical quantities like pressure
gradients and the mass.
\end{abstract}
\section{Introduction}
Most of the stars in the galaxies are the main sequence stars
which evolve by burning lighter elements into heavier nuclei.
Stars massive than $\sim10M_\odot$ explode into supernova leaving
the core behind which then collapses to form a compact object. The
cores are supported by the degenerate pressure of its constituent
particles and possess the densities of the relativistic scales
i.e. $R_s=2GM/rc^2\sim1$ (which is the Schwarzschild radius of the
star) where $M$ is the mass and $r$ is radius of the compact
object. At these densities, relativistic effects dominate and the
physical quantities like gradients of pressure and mass are
determined by the Tolman-Oppenheimer-Volkoff (TOV) equations for
an isotropic and homogeneous compact star. In these stars, the
matter is found in stable ground state where the quarks are
confined inside the hadrons. It is suggested that the quarks if
de-confined into individual $u$, $d$ and $s$ quarks can also yield
a stable ground state of matter, which is called `strange matter'
\cite{witten}. Stars composed of mostly strange matter are
therefore termed as `strange stars' \cite{depaolis} (see
\cite{xu,glen} for review on this topic). Generally, superdense
stars with mass to size ratio exceeding 0.3 are expected to be
composed of strange matter \cite{tike}. The prime motivation for
the existence of strange stars was to explain the exotic phenomena
of gamma ray bursts and soft gamma ray repeaters
\cite{cheng,cheng1}. Now with the observations of the Rossi X-ray
Timing Explorer, it is convincingly shown that astrophysical
source SAX J1808.4-3658 is more likely a strange star \cite{li}.
The transition from the normal hadronic to the strange matter
occurs at sufficiently high densities or corresponding low
temperatures as $T\propto V\propto\rho^{-1}$. These conditions can
mostly be found inside the cores of fast rotating pulsars, P-stars
(composed of $u$ and $d$ quarks and are in $\beta$ equilibrium
with electrons) or magnetic field powered magnetars. The density
distribution inside these stars need not be isotropic and
homogeneous (as proposed in the TOV model) if strange matter truly
exists, then stars composed of entirely strange matter can also be
found. Recently several authors have studied compact stars with
anisotropic matter distribution \cite{prad,bohm,dev,kg,yav}. We
provide star models admitting conformal motion with different
anisotropy.

To search the natural relation between geometry and matter through
the Einstein's Equations, it is useful to use inheritance symmetry.
The well known inheritance symmetry is the symmetry under conformal
killing vectors (CKV) i.e.
\begin{equation}
L_\xi g_{ik} = \psi g_{ik}, \label{Eq3}
\end{equation}
where $L$ is the Lie derivative operator, $\xi$ is the four vector
along which the derivative is taken and $\psi$ is the conformal
killing vector. Note that if $\psi=0$ then Eq. (1) gives the Killing
vector, if $\psi=constant$ it gives homothetic vector and if
$\psi=\psi(\textbf{x},t)$ then it yields conformal vectors. Thus CKV
provides a deeper insight into the spacetime geometry. Moreover, if
the conformal factor $\psi=0$, it implies that the underlying
spacetime is conformally flat which further implies that the Weyl
tensor also vanishes. These conformally flat spacetimes represent
gravitational fields without the source of matter producing these
fields.

The plan of the paper is as follows: In the second section, we model
our gravitational system and formulate the field equations. Next, we
shall solve these field equations using different ansatz of our
parameters. Then we present the graphical representation of our
results. Finally we conclude with the discussion of our results.

\section{The model}

The static spherically symmetric spacetime (in geometrical units
$G=1=c$ here and onwards) is taken as
\begin{equation}
ds^2=   e^{\nu(r)} dt^2-e^{\lambda(r)}
dr^2-r^2(d\theta^2+\sin^2\theta d\phi^2). \label{Eq3}
\end{equation}
The Einstein field equations (EFE) for the above metric are
\begin{eqnarray}
e^{-\lambda} \left[\frac{\lambda^\prime}{r} - \frac{1}{r^2}
\right]+\frac{1}{r^2}&=& 8\pi \rho,\\
e^{-\lambda}
\left[\frac{\nu^\prime}{r}+\frac{1}{r^2}\right]-\frac{1}{r^2}&=&
8\pi p_r,\\
\frac{1}{2} e^{-\lambda} \left[\frac{1}{2}(\nu^\prime)^2+
\nu^{\prime\prime} -\frac{1}{2}\lambda^\prime\nu^\prime +
\frac{1}{r}({\nu^\prime- \lambda^\prime})\right] &=&8\pi p_t.
\end{eqnarray}
Here $\rho$ is the energy density while $p_r$ and $p_t$ are the
radial and transverse pressure densities of the fluid. The conformal
killing equation (1) becomes
\begin{equation}
L_\xi g_{ik} =\xi_{i;k}+ \xi_{k;i} = \psi g_{ik}. \label{Eq3}
\end{equation}
The above equations give the following equations as
\begin{eqnarray}
\xi^1 \nu^\prime &=&\psi,\\
\xi^4  &=& C_1  ,\\
\xi^1 & =& \frac{\psi r}{2}, \\
\xi^1 \lambda ^\prime + 2 \xi^1 _{,1} &=&\psi.
\end{eqnarray}
Integration of Eqs. (7-10) yield
\begin{eqnarray}
e^\nu  &=&C_2^2 r^2 , \\   \label{Eq3}
 e^\lambda  &=& \left(\frac {C_3}
{\psi}\right)^2 ,  \\  \label{Eq3} \xi^i &=& C_1 \delta_4^i +
\left(\frac{\psi r}{2}\right)\delta_1^i ,\label{Eq3}
\end{eqnarray}
where $C_i$, $i=1,2,3$ are constants of integration. Making use of
Eqs. (11-13) in (3-5), we can write
\begin{eqnarray}
\frac{1}{r^2}\left[1 - \frac{\psi^2}{C_3^2}
\right]-\frac{2\psi\psi^\prime}{rC_3^2}&=& 8\pi \rho,\\
\frac{1}{r^2}\left[1 - \frac{3\psi^2}{C_3^2}
\right]&=& - 8\pi p_r,\\
\left[\frac{\psi^2}{C_3^2r^2}
\right]+\frac{2\psi\psi^\prime}{rC_3^2} &=&8\pi p_t.
\end{eqnarray}
Thus Eqs. (14-16) represent the EFE in terms of the conformal
factor $\psi$ \textbf{\cite{prad,bohm,mak}.} It is generally
assumed that stars are the spherically symmetric anisotropic
fluid, so one has to use the equations (2) - (6)  for the initial
consideration. Since we consider anisotropic star model admitting
conformal motion , so all the parameters $ e^{\nu} $, $e^\lambda
$, $ \rho $, $p_r$, $p_t$  could be found in terms of conformal
factor $ \psi $ . In other words, unless one knows the exact form
of $ \psi $, one could not say any things ( i.e. all the physical
properties of the anisotropic fluid ). Our model is new and
interesting in the sense that we are the first authors who
consider anisotropic stars with different form of anisotropy
admitting conformal motion. The assumption of different
anisotropy leads different differential equations of conformal
factor $ \psi $ i.e $ \psi $ is constrained by the role of
anisotropy. Recent observations on highly compact objects like X
ray pulsar Her X - 1 , etc support the existence of anisotropic
stars. But it is still unknown the stars maintain what type of
anisotropy.  Since, conformal killing vector provides a deeper in
sight in to the space time geometry, so we propose anisotropic
star models with different anisotropy admitting conformal motion.
In the next section, we shall solve these conformal field
equations using different ansatz of the surface tension
$p_t-p_r$. The surface tension actually determines the anisotropy
in the stellar model.

\section{Solution of Conformal EFE}
Now we consider three different cases to get exact analytical
solutions. Since the metric coefficient $g_{tt}$ is independent of
conformal factor $\psi$, so all the cases $g_{tt}$ assumes the
same form.
\begin{figure}
    \centering
        \includegraphics[scale=.4]{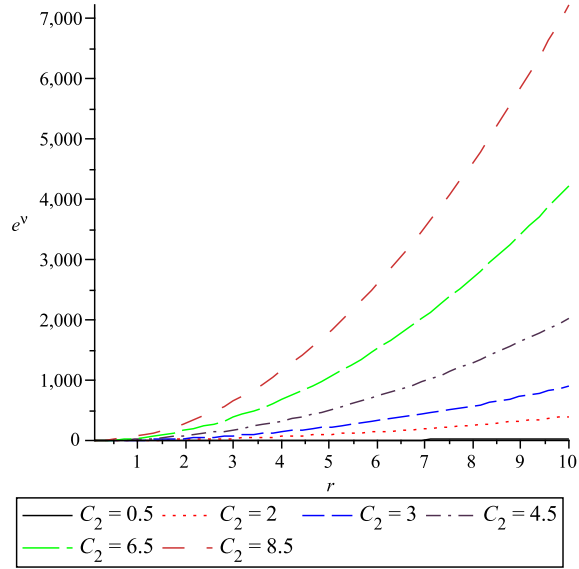}
    \caption{Plot for the variation of $e^\nu$  vs r( km ). }
    \label{}
\end{figure}
\subsection{$p_t = n p_r$, $0<n<1$.}
The first case deals when $p_r$ and $p_t$ are linearly proportional
to each other. It is assumed that $p_t$ is smaller then $p_r$. By
using above anisotropic relation, Eqs. (15) and (16) together will
give a non-homogeneous and non-linear differential equation as
\begin{equation}
 \frac{2 \psi^\prime r}{\psi}
+ \frac{n C_3^2}{\psi^2}= ( 3n - 1).
\end{equation}
Solving this equation, we get
\begin{equation}
\psi = \sqrt{\left[ (rB)^{3n-1} + \frac {n C_3^2}{3n-1}\right]},
\end{equation}
where $B$ is the constant of integration and will be determined
later. Hence we get the exact analytical form all the parameters as
\begin{eqnarray}
p_t  &=&  \frac{ n \left[ C_3^2 + 3(3n-1)(rB)^{(3n-1)}\right]}{8\pi
C_3^2(3n-1)r^2},\\
p_r & =&  \frac{  \left[ C_3^2 + 3(3n-1)(rB)^{(3n-1)}\right]}{8\pi
C_3^2(3n-1)r^2},\\
e^\nu  &=&C_2^2 r^2\\
e^\lambda & =& \frac {C_3^2} {\left[ (rB)^{3n-1} + \frac {n
C_3^2}{3n-1}\right]},\\
\rho  &=&\frac{1}{8 \pi r^2}-   \frac {\left[ (rB)^{3n-1} + \frac {n
C_3^2}{3n-1}\right]} { 8 \pi C_3^2 r^2} - \frac {\left[ B(3n-1)
(rB)^{3n-2} \right]} {8 \pi C_3^2 r}.
\end{eqnarray}
Differentiation of Eq. (20) with respect to $r$ yields the pressure
gradient
\begin{equation}
 8 \pi \frac{dp_r}{dr}  = \frac{  \left[  3B(3n-1)^2(rB)^{(3n-2)}\right]}{ C_3^2(3n-1) r^2}
 - \frac{  2 \left[ C_3^2 + 3(3n-1)(rB)^{(3n-1)}\right]}{C_3^2(3n-1)
 r^3}.
\end{equation}
Here, one can note that the pressure gradient is a decreasing
function of $r$ ( see figure 9). In order to get some physically
meaningful solution, we need to determine the constant $B$. Since
vanishing of radial pressure at the boundary is a consequence of
the junction condition \textbf{\cite{chat}}, so one can use
\begin{equation}
p_r(r=R)=0.
\end{equation}
to find the constant B.

 Notice that Eq. (25) is motivated due to
the fact that pressure gradient is a decreasing function of $r$.
Thus using Eq. (20) in (25) yields
\begin{equation}
B=\frac{1}{R}\left[\frac{C^2_3}{3(1-3n)}\right]^{\frac{1}{3n-1}}.
\end{equation}
Using Eq. (26) in (18), we get
\begin{equation}
\psi=\sqrt{\frac{nC^2_3}{1-3n}+\left[\frac{3^{\frac{1}{1-3n}}
(\frac{C^2_3}{1-3n})^{\frac{1}{3n-1}}r}{R} \right]^{3n-1}}.
\end{equation}
The mass function is
\begin{equation} m(r) = \int_0^r 4\pi \rho r^2 dr = \frac{r(2n-1)}{2(3n-1)} -
\frac{ (rB)^{3n}}{2 BC_3^2}.
\end{equation}
Thus using Eq. (26) in (28) we obtain
\begin{equation}
 m(r) = \frac{r(2n-1)}{2(3n-1)} - \frac{R
r^{3n}}{2
C_3^2}\left(\frac{C_3^2}{3-9n}\right)^{\frac{1}{1-3n}}\left[\frac{\left(\frac{C_3^2}{3-9n}\right)^{\frac{1}{3n-1}}r}{R}\right]^{3n}.
\end{equation}
Since for $ n \geq \frac{1}{3}$, the solutions are inconsistent,
so we neglect the cases $ n \geq \frac{1}{3}$. We also observe
that $r=0$ gives a singularity. However, the solutions are valid
for some radius $r>0$. Now, we calculate   the subluminal sound
speed , $ \mid v_s^2 \mid  =  \mid \frac { dp}{d\rho}\mid  $ \\as
\[ \mid v_s^2 \mid  = \frac{2[C_3^2 +3(3n-1)^2(rB)^{3n-1} ] -
3(3n-1)^2(rB)^{3n-1}}{3n(3n-1)^2(rB)^{3n-1}+2(3n-1)C_3^2-2(3n-1)(rB)^{3n-1}-2nC_3^2-2(3n-1)^2(rB)^{3n-1}}\]
The above expression should be less than one depending on the
parameters. Thus sound speed does not exceed that of light as
fulfillment of causality condition.
\begin{figure}
    \centering
        \includegraphics[scale=.33]{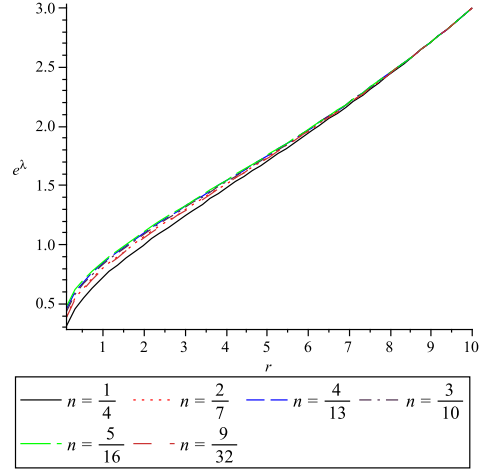}
    \caption{Plot for the variation of $e^\lambda$  vs r( km ). Here we assume the radius of the star $R= 10$  km
    and $C_3 =1$.}
    \label{}
\end{figure}
\begin{figure}
    \centering
        \includegraphics[scale=.33]{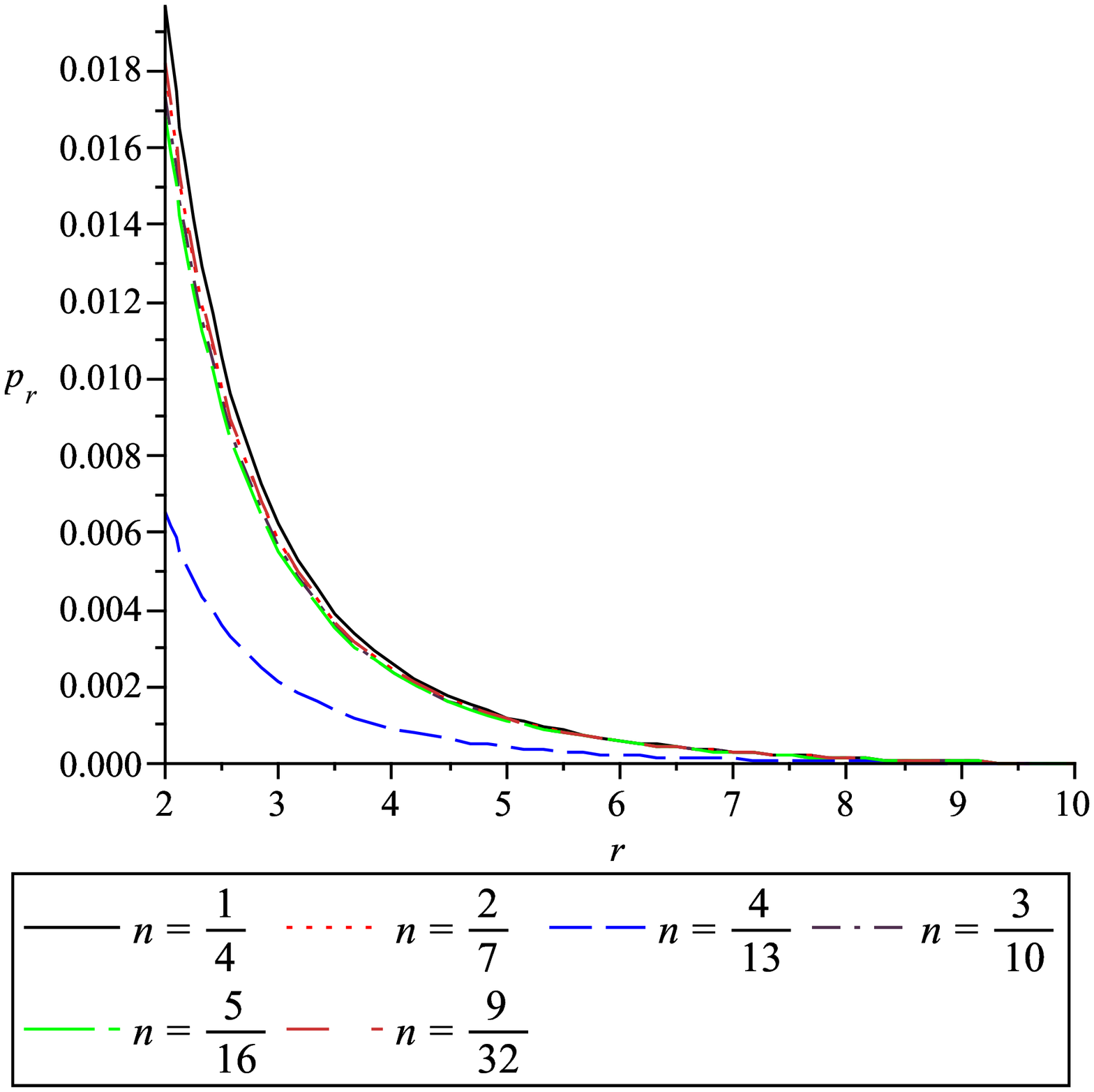}
    \caption{Plot for the variation of radial pressure $p_r$  vs r( km ). Here we assume the radius of the star $R= 10$  km
    and $C_3 =1$.}
    \label{}
\end{figure}
\begin{figure}
    \centering
        \includegraphics[scale=.33]{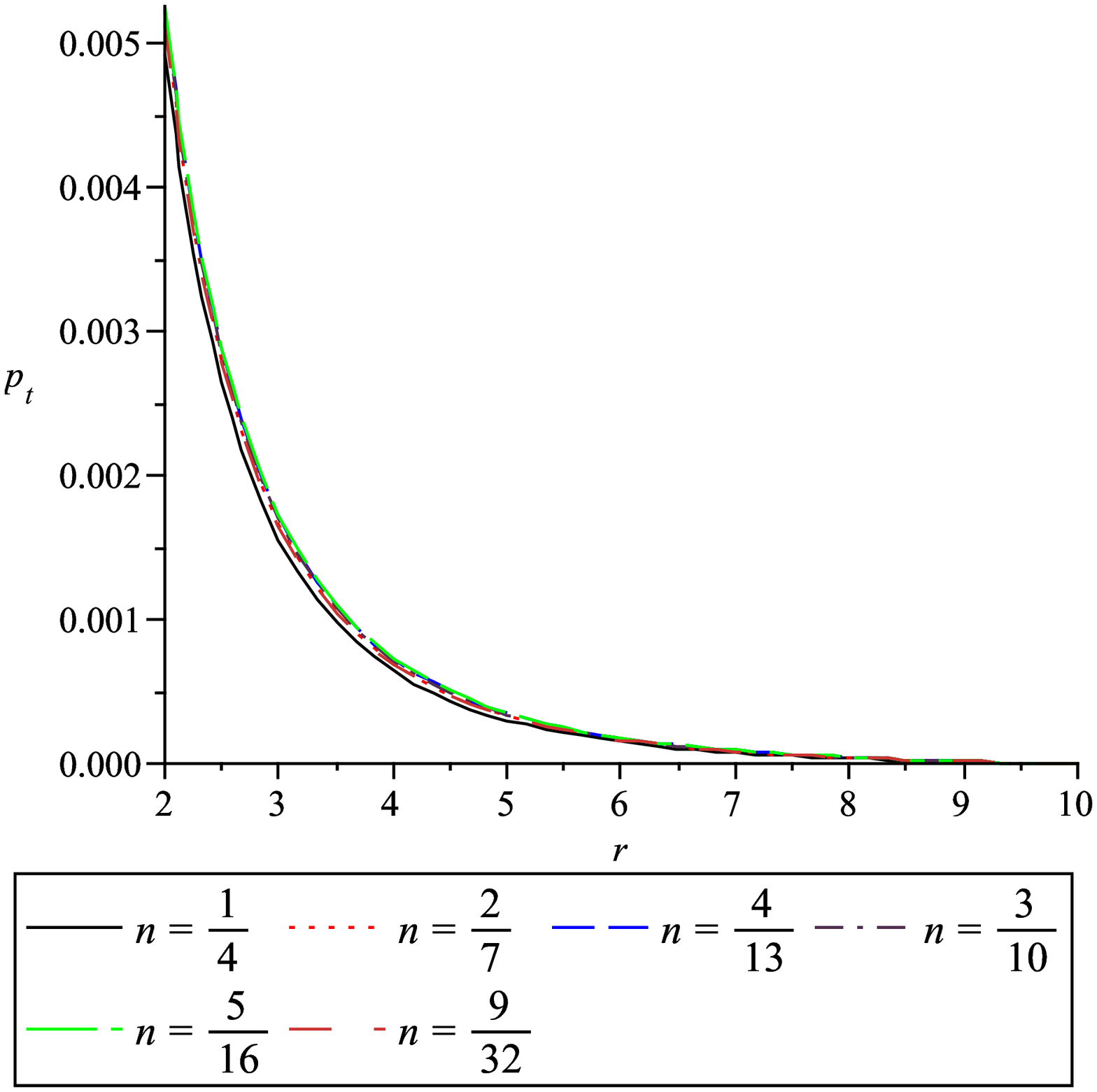}
    \caption{Plot for the variation of transverse  pressure $p_t$  vs r( km ). Here we assume the radius of the star $R= 10$  km
    and $C_3 =1$. }
    \label{}
\end{figure}
\begin{figure}
    \centering
        \includegraphics[scale=.33]{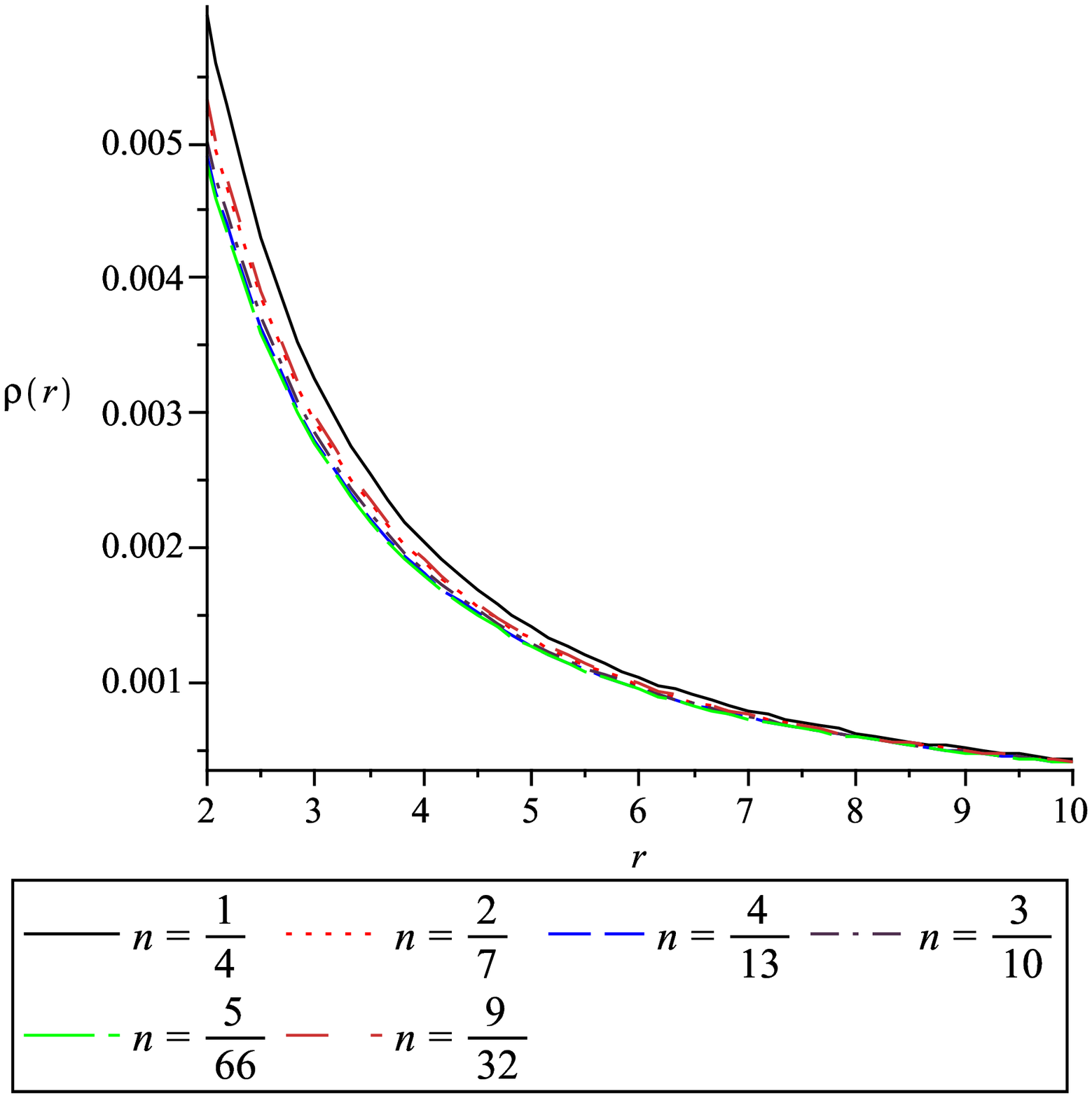}
    \caption{Plot for the variation of energy density $\rho$  vs r( km ). Here we assume the radius of the star $R= 10$  km
    and $C_3 =1$. }
    \label{}
\end{figure}
\begin{figure}
    \centering
        \includegraphics[scale=.33]{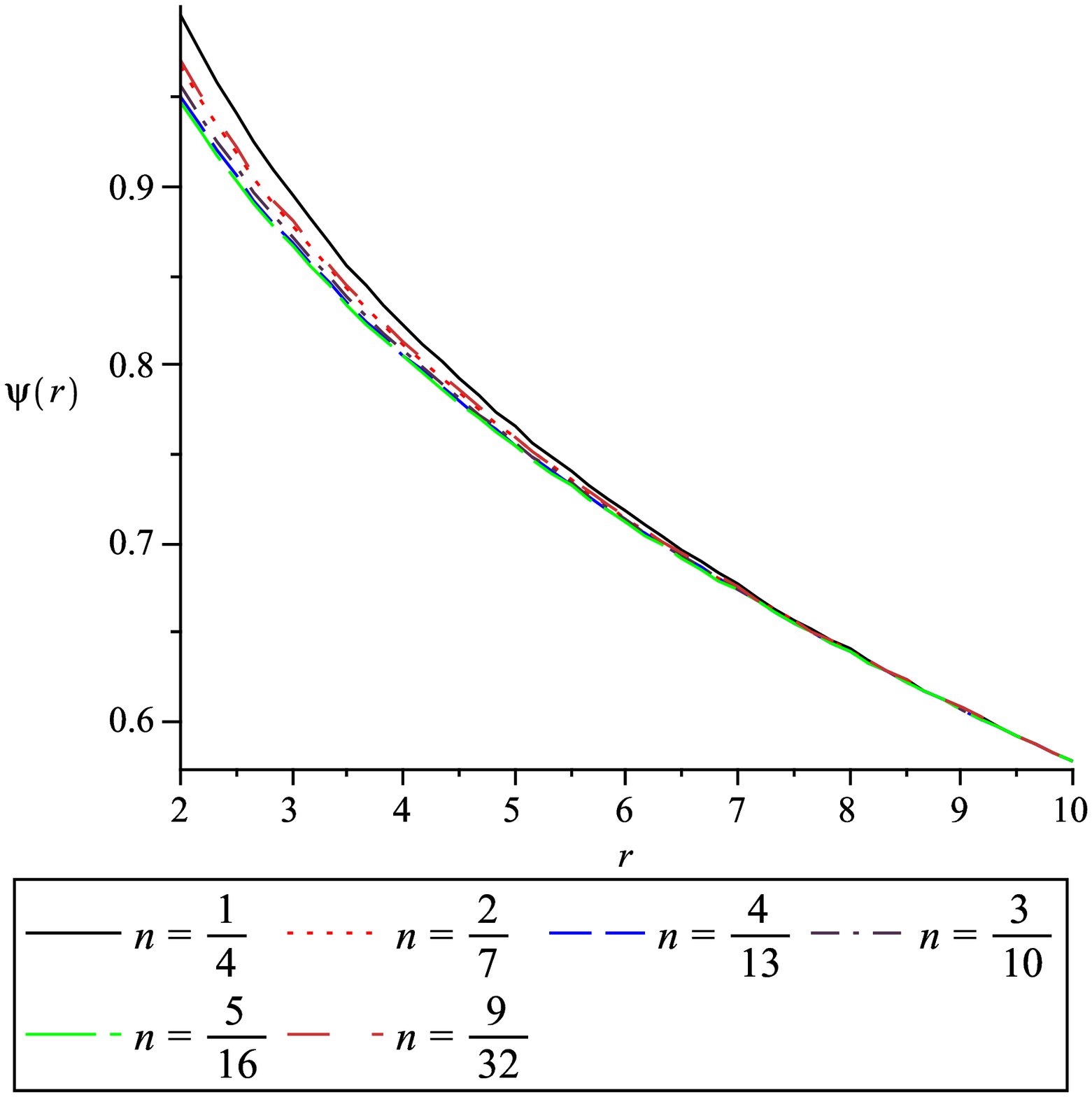}
    \caption{Plot for the variation of the conformal factor $\psi$  vs r( km ). Here we assume the radius of the star $R= 10$  km
    and $C_3 =1$. }
    \label{}
\end{figure}
\begin{figure}
    \centering
        \includegraphics[scale=.33]{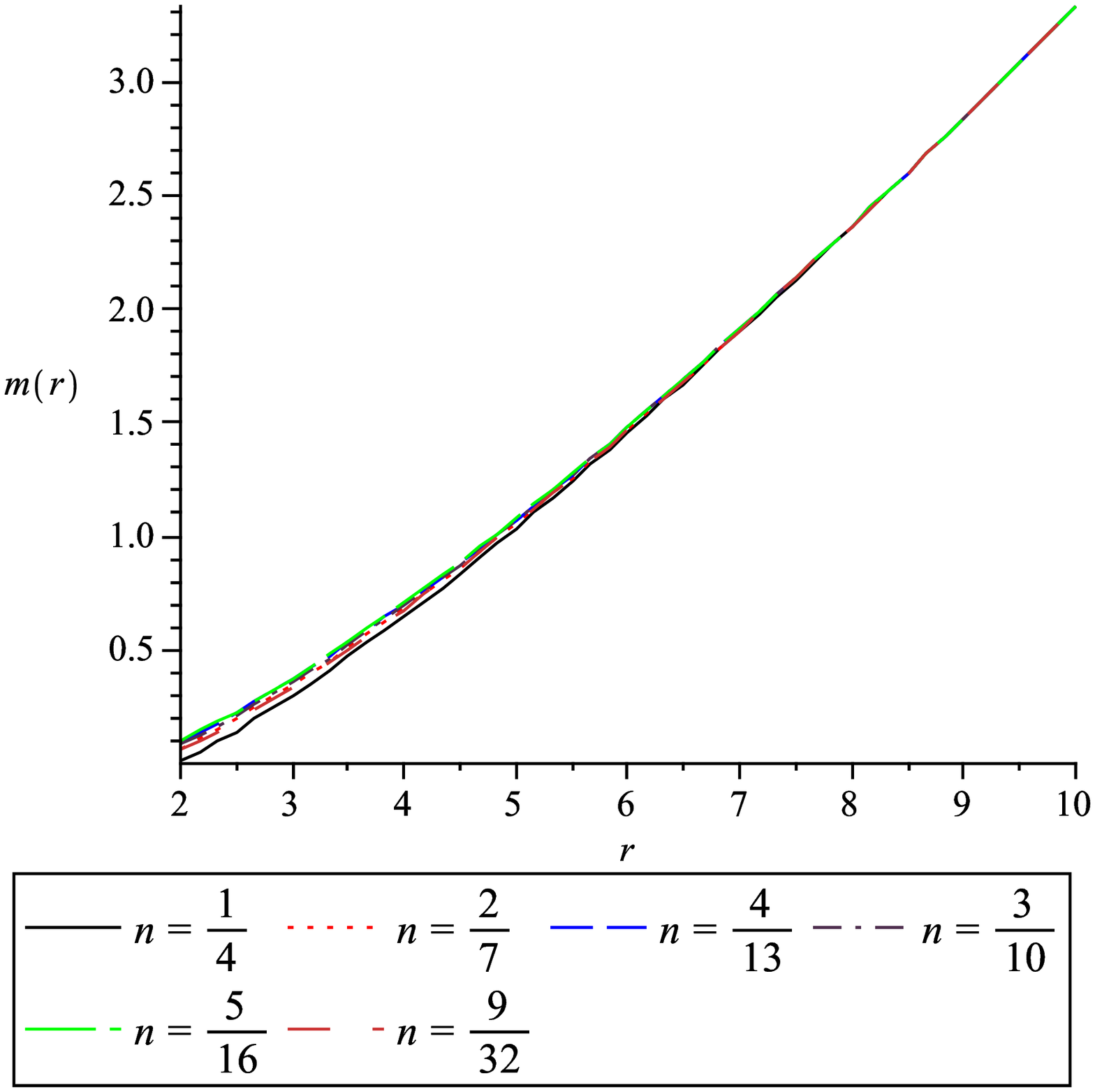}
    \caption{Plot for the variation of mass $m(r)$  vs r( km ). Here we assume the radius of the star $R= 10$  km
    and $C_3 =1$. }
    \label{}
\end{figure}
\begin{figure}
    \centering
        \includegraphics[scale=.33]{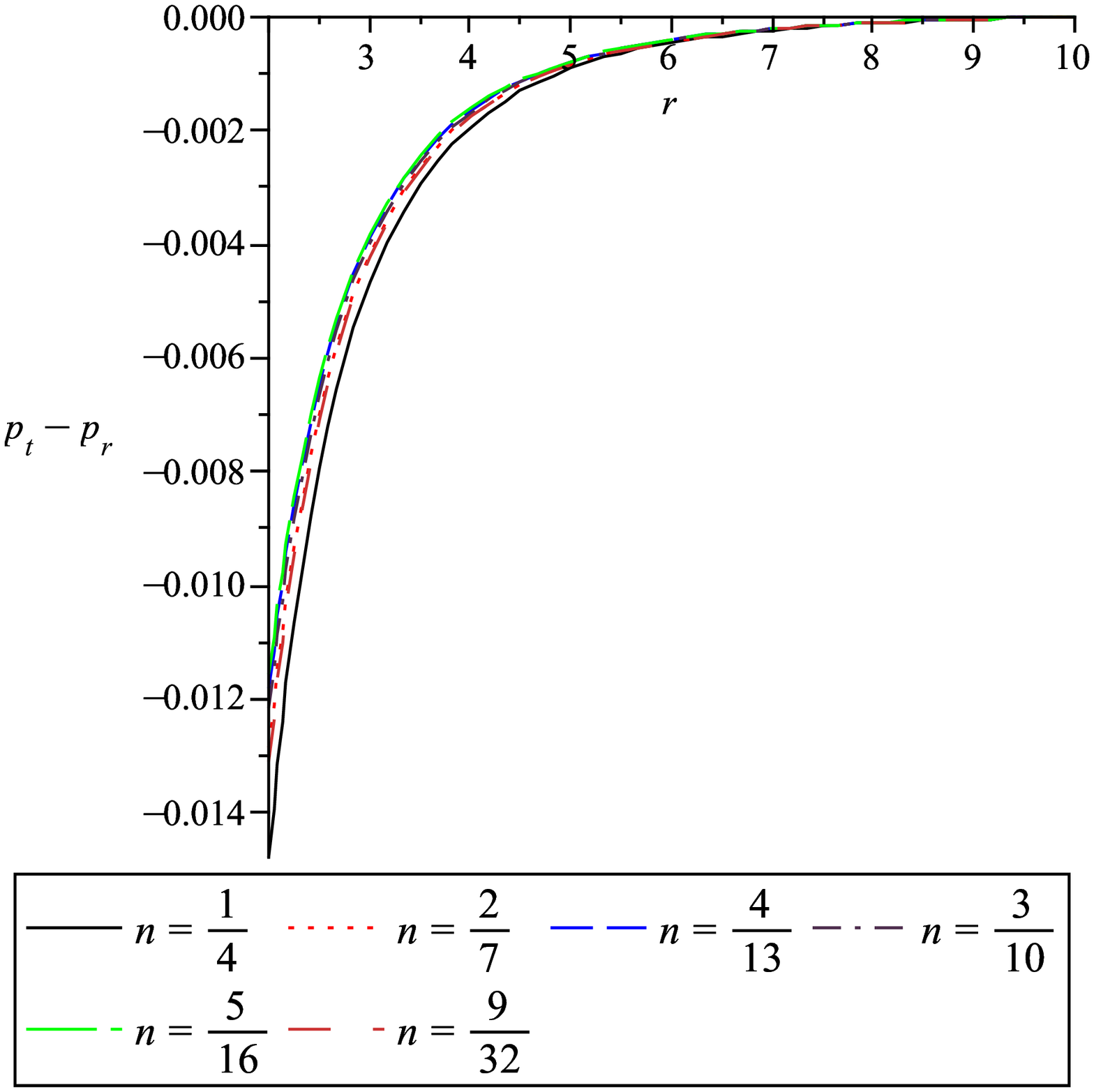}
    \caption{Plot for the variation of  the  pressure anisotropy  $p_t - p_r$  vs r( km ). Here we assume the radius of the star $R= 10$  km
    and $C_3 =1$. }
    \label{}
\end{figure}
\begin{figure}
    \centering
        \includegraphics[scale=.33]{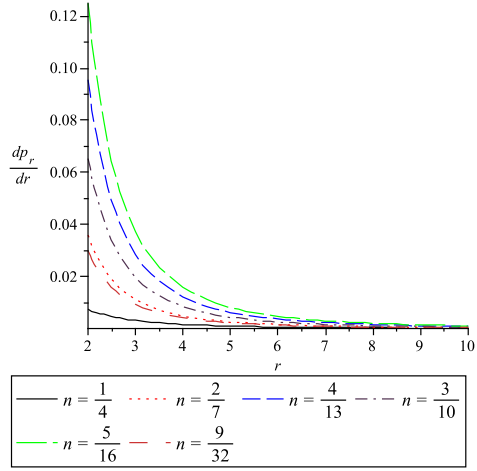}
    \caption{ Plot for the variation of  the  gradient of radial pressure   $\frac{dp_r}{dr}$  vs r( km ). Here we assume the radius of the star $R= 10$  km
    and $C_3 =1$. }
    \label{}
\end{figure}

\pagebreak

\subsection{$p_t - p_r = \frac{1}{8 \pi}( \frac{c_1}{r^2} + c_2)$, where $c_1$ and $c_2$ are arbitrary constants.}

 By using above anisotropic relation \textbf{[7]}, we get from Eqs.
 (15) and (16)
\begin{equation}
 \frac{2\psi  \psi^\prime }{C_3^2 r}
- \frac {2 \psi^2}{ C_3^2 r^2} = \frac{( c_1 - 1)}{r^2} + c_2.
\end{equation}
Solving this equation, we get
\begin{equation}
\psi = \sqrt{ - \frac{A}{2} + B r^2 \ln r + K_1 r^2},
\end{equation}
where $ A = C_3^2 ( c_1-1)$, $B = C_3^2 c_2 $ and $K_1$ is an
integration constant. Thus we get the exact analytical form of all
the parameters as
\begin{eqnarray}
p_t  &=&  \frac{  \left[ -A + 4Br^2 \ln r + 2B+ 4K_1
r^2\right]}{16\pi C_3^2r^2},\\
p_r  &=&  \frac{ 3( - \frac{A}{2} + B r^2 \ln r + K_1 r^2) }{8\pi
C_3^2
r^2}- \frac{1}{8 \pi r^2},\\
e^\nu & =&C_2^2 r^2,\\
e^\lambda  &=& \frac {C_3^2} {\left[  - \frac{A}{2} + B r^2 \ln r +
K_1 r^2
\right]},\\
\rho  &=& \frac {\left[   A - 6 B r^2 \ln r -2B r^2 - 6 K_1 r^2 +
2C_3^2 \right]}{ 16 \pi C_3^2 r^2}.\\
\end{eqnarray}
The mass function is
\begin{equation}
 m(r) = \frac{\left[Ar - \frac{2B r^3}{3} - 2 K_1 r^3 + 2C_3^2 r -
3B r^3 ( \ln r- \frac{1}{3})\right]}{4 C_3^2}.
\end{equation}
The pressure gradient is given by
\begin{equation}
  \frac{dp_r}{dr}  = - \frac{ 3( - \frac{A}{2} + B r^2 \ln r + K_1 r^2) }{4 \pi C_3^2
 r^3}+ \frac{ 3(  2B r  \ln r +  Br + 2 K_1 r ) }{8 \pi C_3^2
 r^2} + \frac{1}{4 \pi r^3}.
\end{equation}
Applying the similar procedure as done in the previous section,
$p_r(r=R)=0$ gives
\begin{equation}
K_1=\frac{1}{R^2}\left[\frac{C^2_3}{3}+\frac{A}{2}-BR^2\ln
R\right]=\frac{C_3^2}{R^2}\left[ \frac{c_1}{2}-\frac{1}{6}-c_2R^2\ln
R \right].
\end{equation}

As before,  we calculate   the subluminal sound speed , $ \mid
v_s^2 \mid  =  \mid \frac { dp}{d\rho}\mid  $ \\as
\[ \mid v_s^2 \mid  = \frac{6( Br^2\ln r +K_1r^2-\frac{A}{2}) -2C_3^2-3r^2(2B\ln r + B+2K_1)}
{r^2(6B\ln r + 5B+6K_1)+( A-6Br^2\ln r -2Br^2-6K_1r^2+2C_3^2)}\]
The above expression should be less than one depending on the
parameters. Thus sound speed does not exceed that of light as
fulfillment of causality condition.

\begin{figure}
    \centering
        \includegraphics[scale=.33]{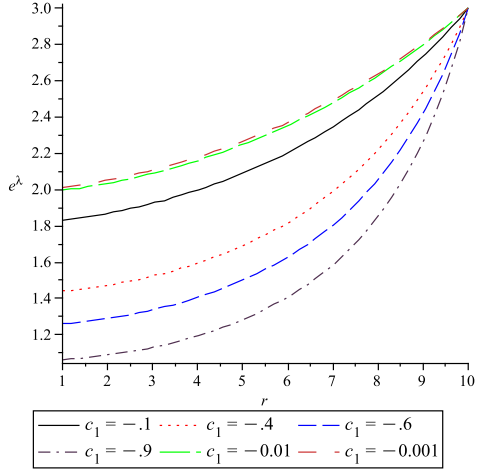}
    \caption{Plot for the variation of $e^\lambda$  vs r( km ). Here we assume the radius of the star $R= 10$  km
    with $c_2 =.001$ and $C_3 =1$. }
    \label{}
\end{figure}
\begin{figure}
    \centering
        \includegraphics[scale=.33]{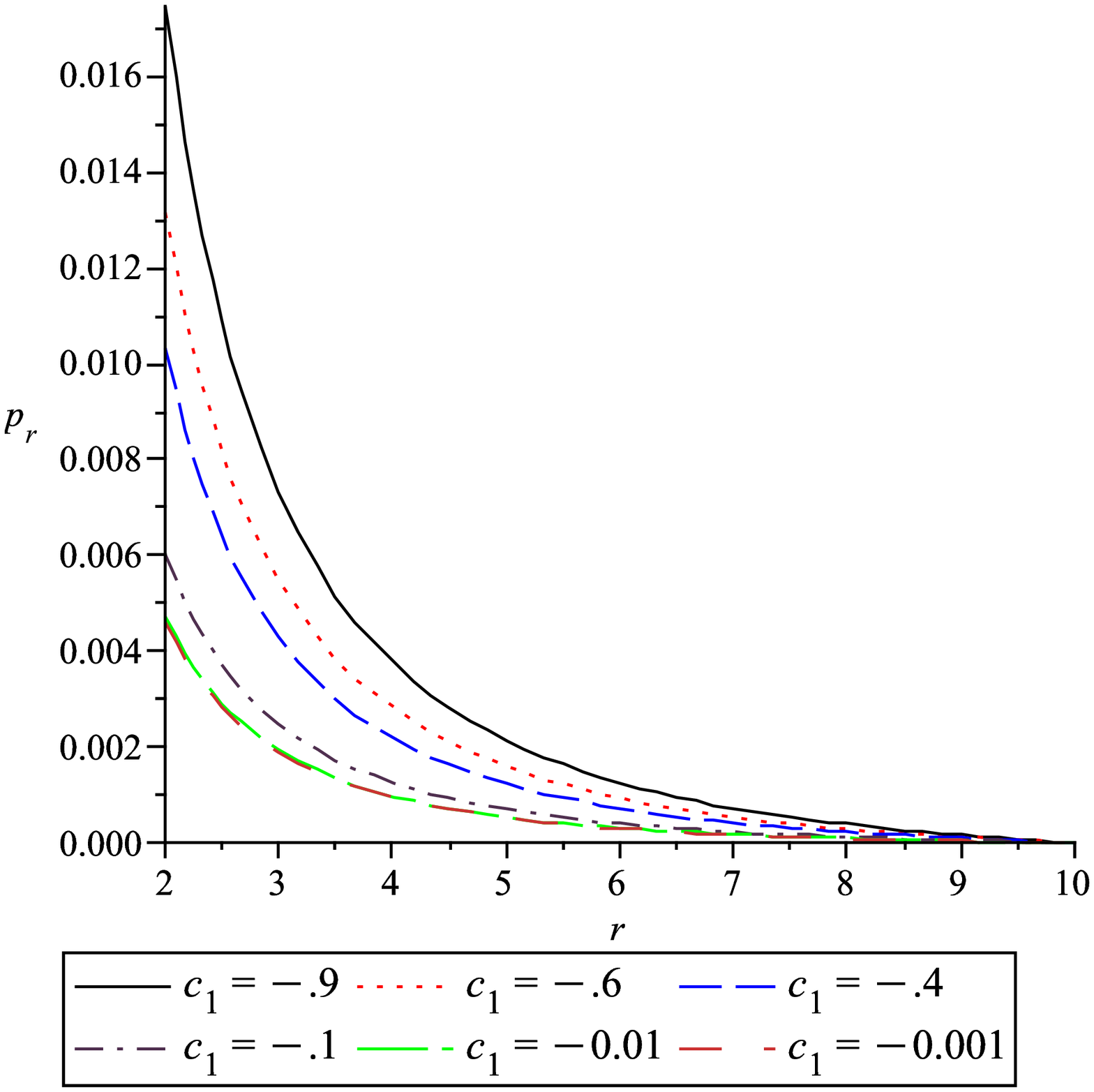}
    \caption{Plot for the variation of radial pressure   $p_r$  vs r( km ). Here we assume the radius of the star $R= 10$  km
    with $c_2 =.001$ and $C_3 =1$. }
    \label{}
\end{figure}
\begin{figure}
    \centering
        \includegraphics[scale=.33]{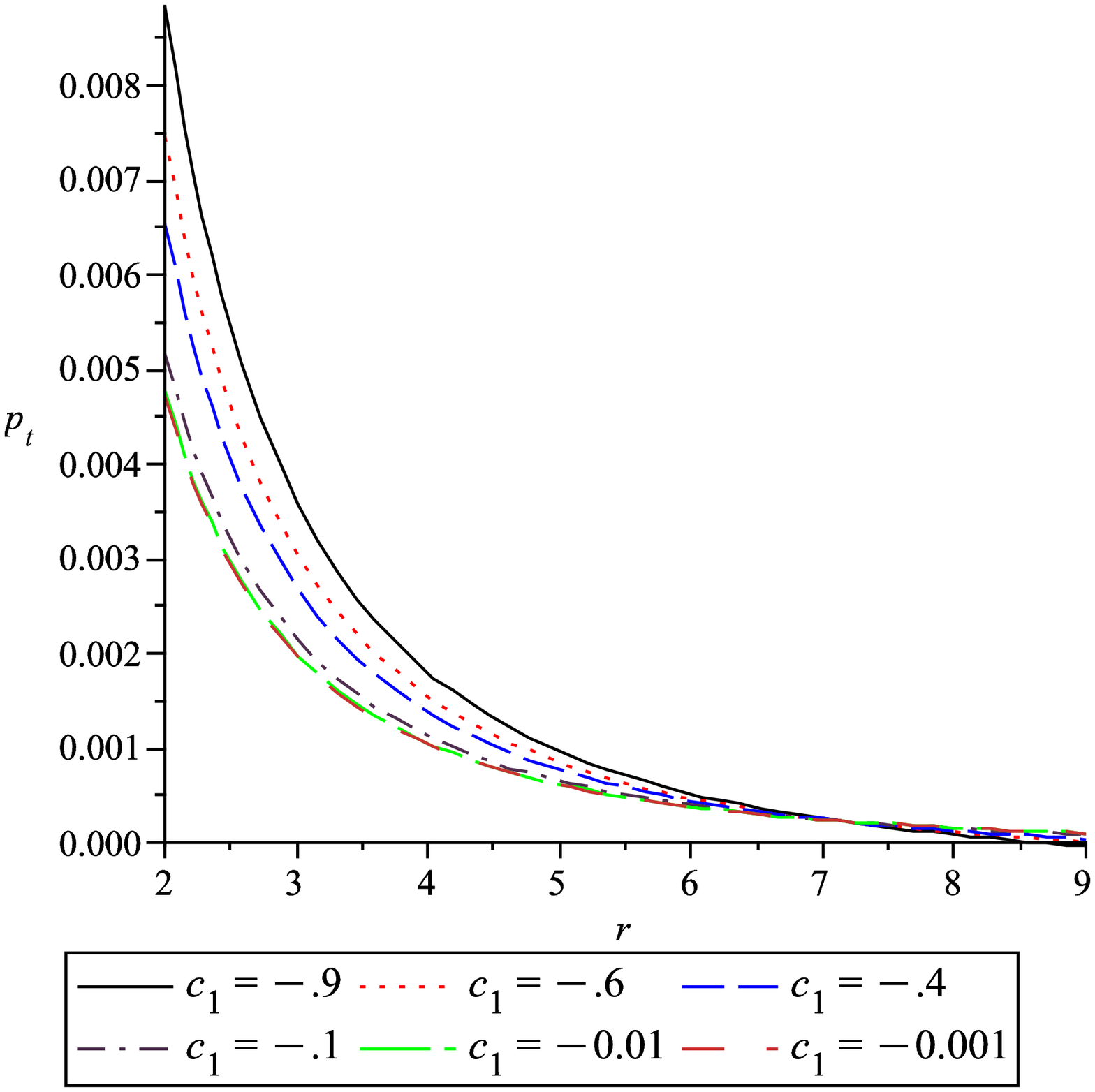}
    \caption{Plot for the variation of transverse pressure   $p_t$  vs r( km ). Here we assume the radius of the star $R= 10$  km
    with $c_2 =.001$ and $C_3 =1$. }
    \label{}
\end{figure}
\begin{figure}
    \centering
        \includegraphics[scale=.33]{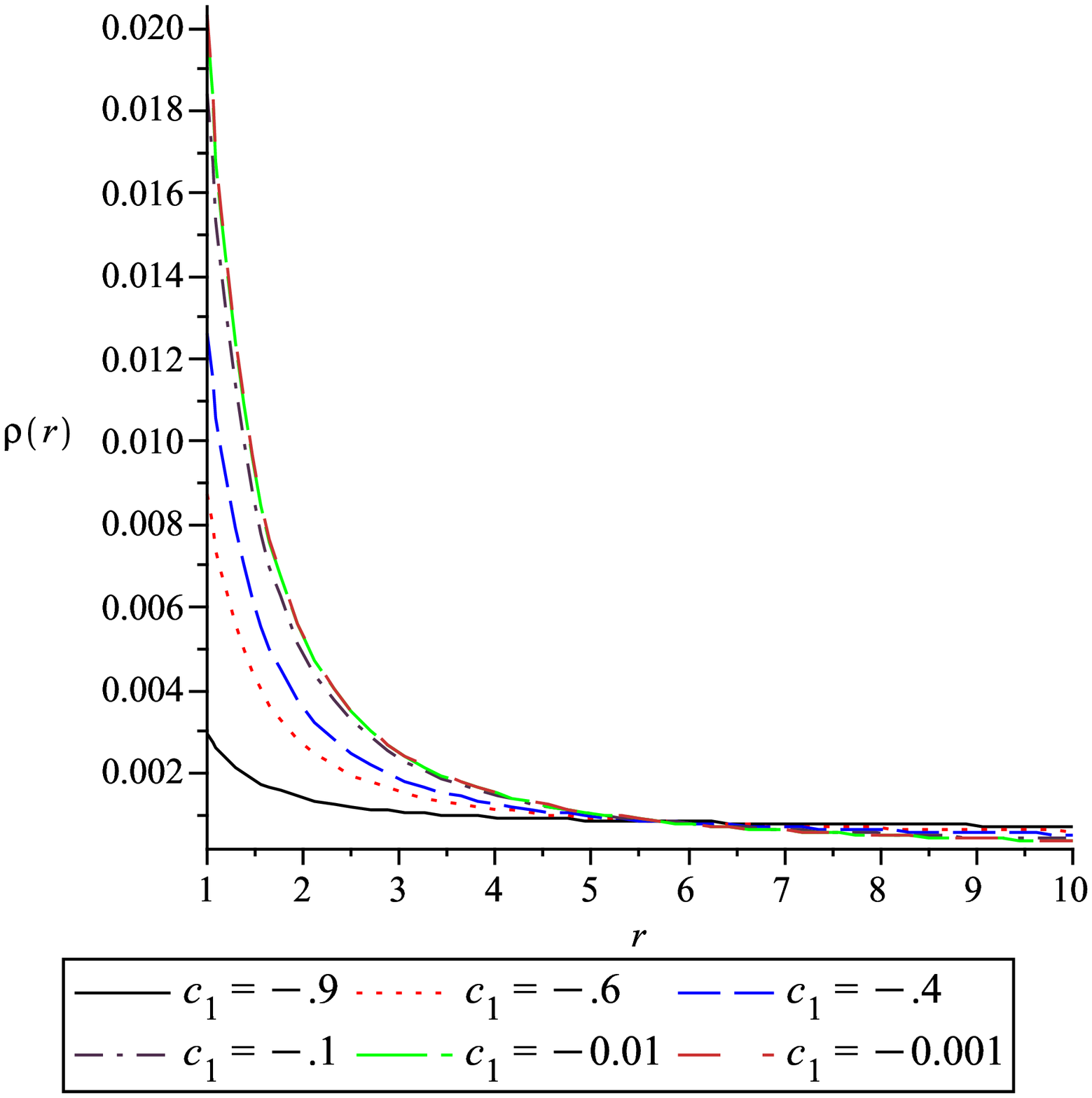}
    \caption{ Plot for the variation of energy density $\rho$  vs r( km ). Here we assume the radius of the star $R= 10$  km
    with $c_2 =.001$ and $C_3 =1$. }
    \label{}
\end{figure}
\begin{figure}
    \centering
        \includegraphics[scale=.33]{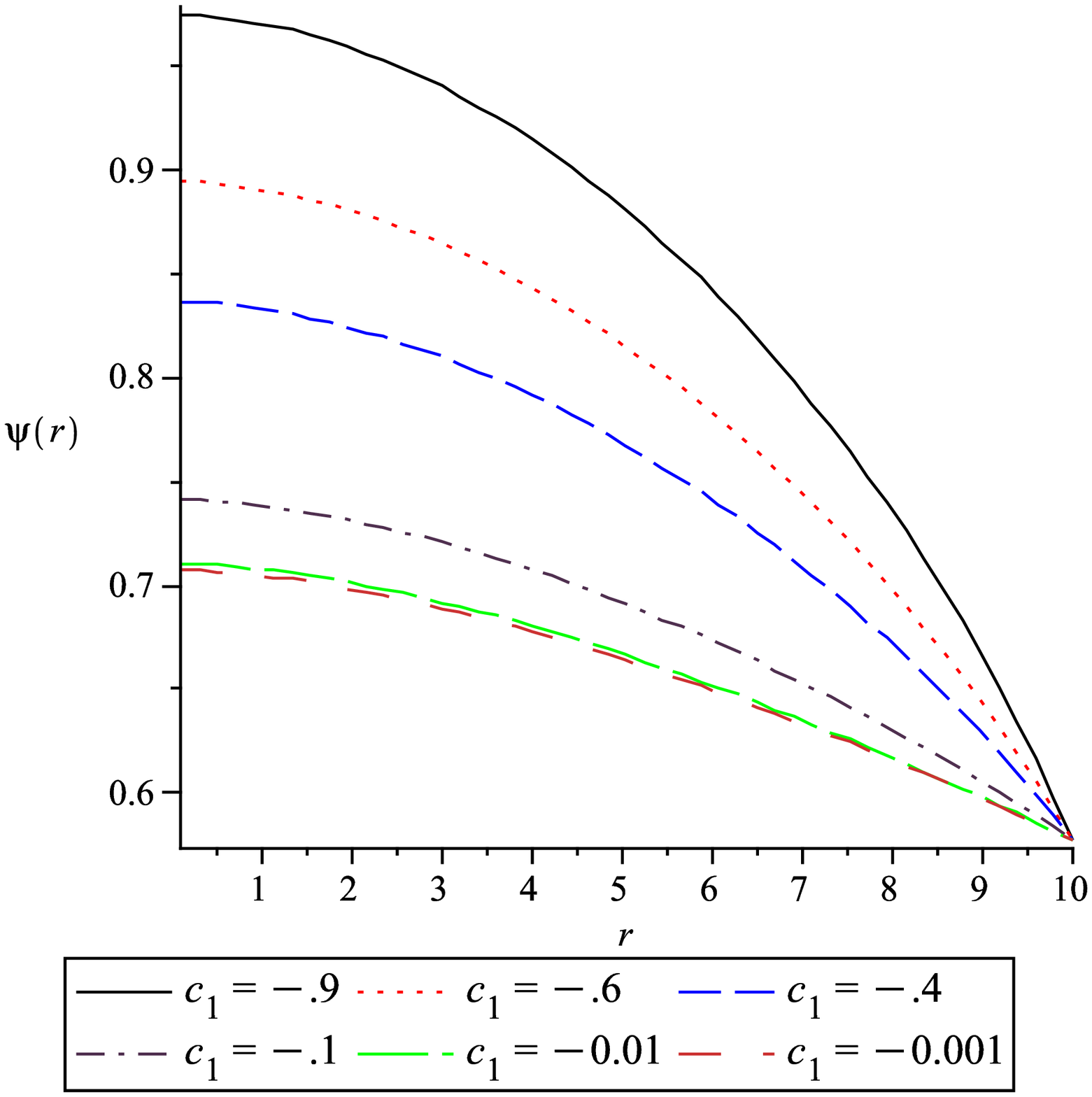}
    \caption{Plot for the variation of the conformal factor $\psi$  vs r( km ). Here we assume the radius of the star $R= 10$  km
    with $c_2 =.001$ and $C_3 =1$.}
    \label{}
\end{figure}
\begin{figure}
    \centering
        \includegraphics[scale=.33]{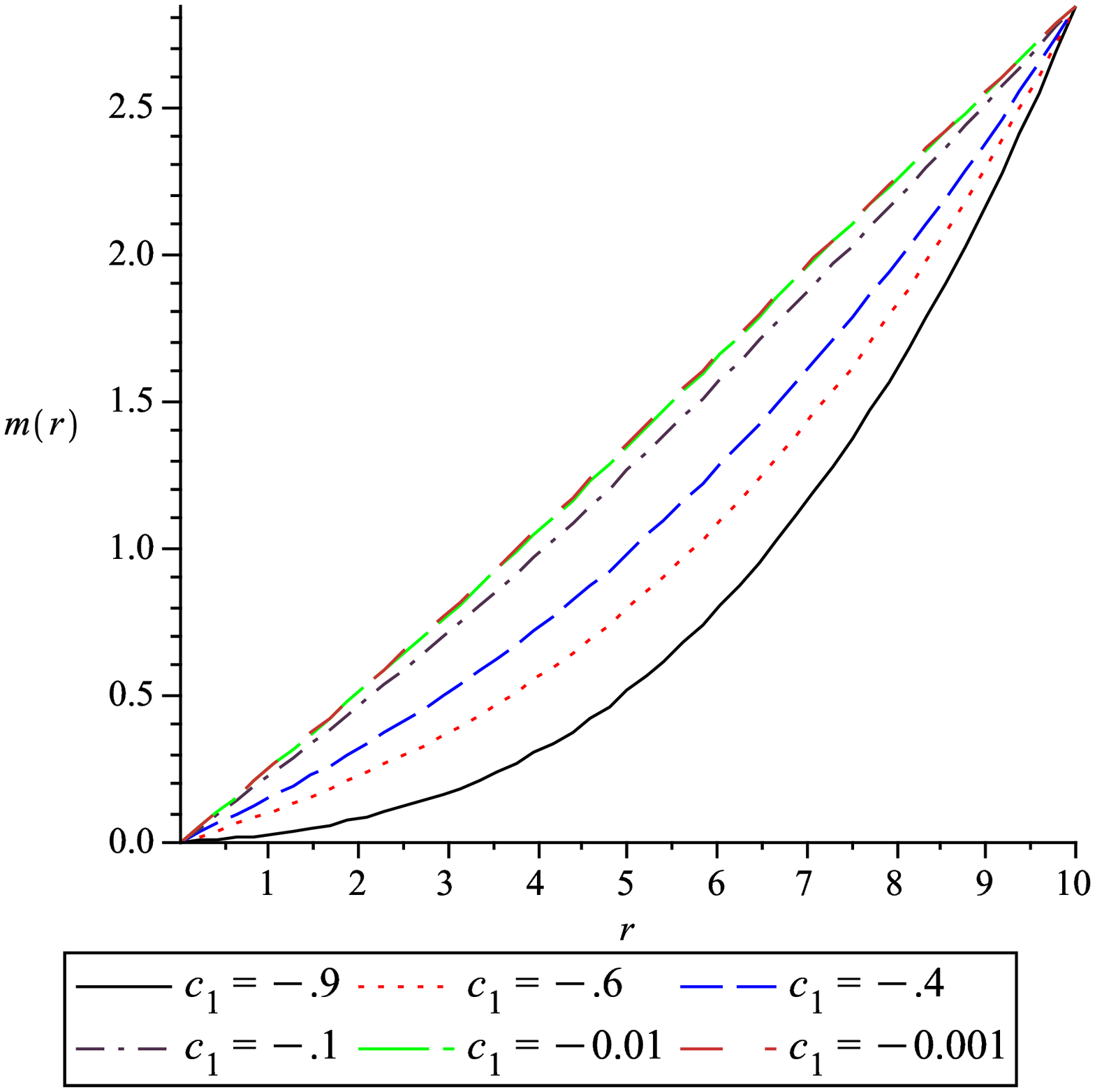}
    \caption{Plot for the variation of mass $m(r)$   vs r( km ). Here we assume the radius of the star $R= 10$  km
    with $c_2 =.001$ and $C_3 =1$. }
    \label{}
\end{figure}
\begin{figure}
    \centering
        \includegraphics[scale=.33]{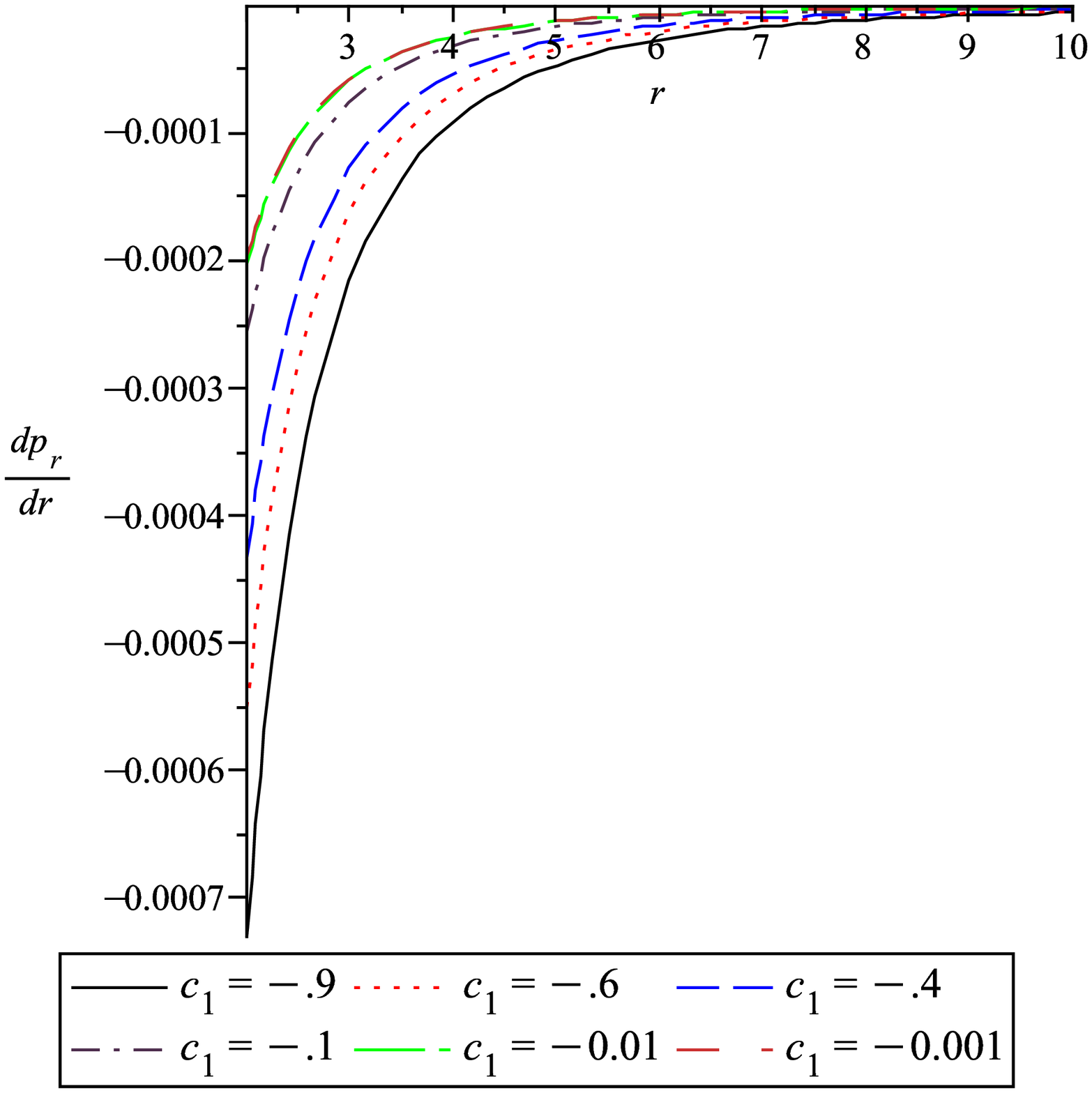}
    \caption{Plot for the variation of  the  gradient of radial pressure   $\frac{dp_r}{dr}$   vs r( km ). Here we assume the radius of the star $R= 10$  km
    with $c_2 =.001$ and $C_3 =1$. }
    \label{}
\end{figure}
\begin{figure}
    \centering
        \includegraphics[scale=.33]{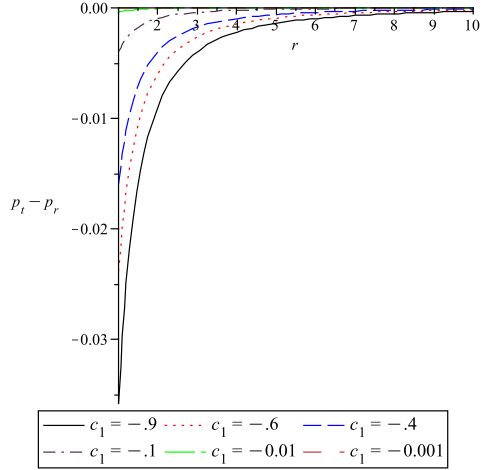}
    \caption{Plot for the variation of the  pressure anisotropy  $p_t - p_r$   vs r( km ). Here we assume the radius of the star $R= 10$  km
    with $c_2 =.001$ and $C_3 =1$. }
    \label{}
\end{figure}

\pagebreak

\section{Conclusion and Discussion}

In the previous section, we have given the pictorial
representation of the parameters involved in the three cases. The
constant of integration is fixed at $C_3=1$ while the radius of
the compact star is chosen to be $R=10$km. All the other physical
parameters like the metric functions $e^\lambda$ and $e^\nu$,
radial pressure $p_r$, transverse pressure $p_t$, surface tension
$p_r-p_t$, mass function $m(r)$, density $\rho(r)$ and the radial
pressure gradient $dp_r/dr$ are plotted against the radial
parameter $r$ in all the cases.\\
In figures 1-9, we have taken different values of $n$ in the range
(0,1) to determine parameters from case-1. We observe that the
radial and transverse pressures gradually decrease as $r$
approaches to the surface $R$. This result is consistent with our
boundary condition of vanishing pressures all at the stellar
surface. It comes as no surprise that the pressure and the mass
profile of the compact stars is all along similar to the normal
hydrogen burning stars. More specifically, the mass profile of the
star shows sharp increase for $n=\frac{5}{16}$ while it is small
for values $n<\frac{5}{16}$. The surface tension $p_t-p_r$ of the
star also decreases along with increasing $r$. Note that the
surface tension varies along negative range due to higher radial
pressure. Thus transverse pressure although exists but is smaller
then the radial pressure along the allowed range of parameter
$r$. The pressure gradient $\nabla_rp_r$ or $dp_r/dr$ also
decreases outward along $r$. For small values of parameter $n$,
the gradient is steeper then for large values.  The density
profile of the compact star falls for increasing $r$. The
conformal factor $\psi(r)$ assumes maximum value at the star's
center while it approaches zero near the stellar surface.
Therefore, the vanishing conformal factor at the star's surface
could be used as an alternative boundary
condition for solving conformal field equations.\\
Similarly, the parameters obtained in case-2 are plotted in
figures 10-17 using different values of parameter $c_1$. The
constant parameter is fixed at $c_2=0.001$. The pressure profile
of the star in this case is similar to the earlier in case-1
having steep slopes with increasing $r$. Also, the density profile
and the conformal factor show convergence for large values of $r$.
The mass profile is steady for $c_1=-0.001$ while it is steepest
for $c_1=-0.9.$ Similarly the declining pressure gradient profile
is observed for large $r$ which is consistent with our boundary
condition Eq.(25).\\
In figures 18-25, we have made a comparison of various stellar
parameters obtained in cases 1 and 2. It is observed that the
metric function $e^\lambda$, conformal factor $\psi$, radial
pressure $p_r$ and the transverse pressure $p_t$ of both cases
converge as $r$ tends to $R$. The mass parameter $m(r)$ is steeper
in the second case then the first one. Therefore the mass will
increase fast radially in the linear model of the surface tension.
Moreover, the surface tension profile of both models goes
asymptotically parallel to each other. The density profile in the
first case goes to zero while it remains constant in the second
case in the asymptotic limit of large $r$.\\
 One can note that any
static spherically symmetric space time admitting conformal motion
suffers $r=0$ singularity. In our model also, we had to tolerate
$r=0$ singularity. Therefore the model does not exist at $r=0$.
However, our solutions are valid for some radius $r>0$. We also
note that there is a singularity in the mass density,  in spite
of, the total mass is finite. So, the models ( case - I and case -
II ) are physically acceptable. In addition to this, one can find
that the matter density and fluid pressure are non negative and
gradient $\frac{dp_r}{dr}$ is decreasing with r. Thus our models (
i.e. interior solutions of the gravitational field equations ) are
fully physically meaningful. As our models suffer $r=0$
singularity, the minimum value of r is taken to be a positive
quantity ( km ). Since radius of the star is taken to 10 km (
this is justified as the radius of some of the observed strange
stars are nearly equal to 10 km ), so one can note that the total
mass of the star is m ( r=10). Our results have been shown in
fig. 7
and fig.15. \\
In our current analysis, we have used an idealization of spherical
symmetry, stationary and static compact star. In general, from the
astrophysical point of view, most of the observed or predicted
compact star candidates are no longer static but rotating about a
unique axis that may be oblique from the magnetic axis of the
star. More exotic stars including magnetars (driven by mostly
magnetic fields) and pulsars (mostly driven by higher angular
momentum) are well known examples of rotating compact stars. In
our forth coming work, we plan to work out a similar analysis
presented here for the fast rotating or ultra-fast rotating
compact stars. We also plan to work on the rotating stars with the
slow rotation approximation as well.

\pagebreak

\begin{figure}
    \centering
        \includegraphics[scale=.33]{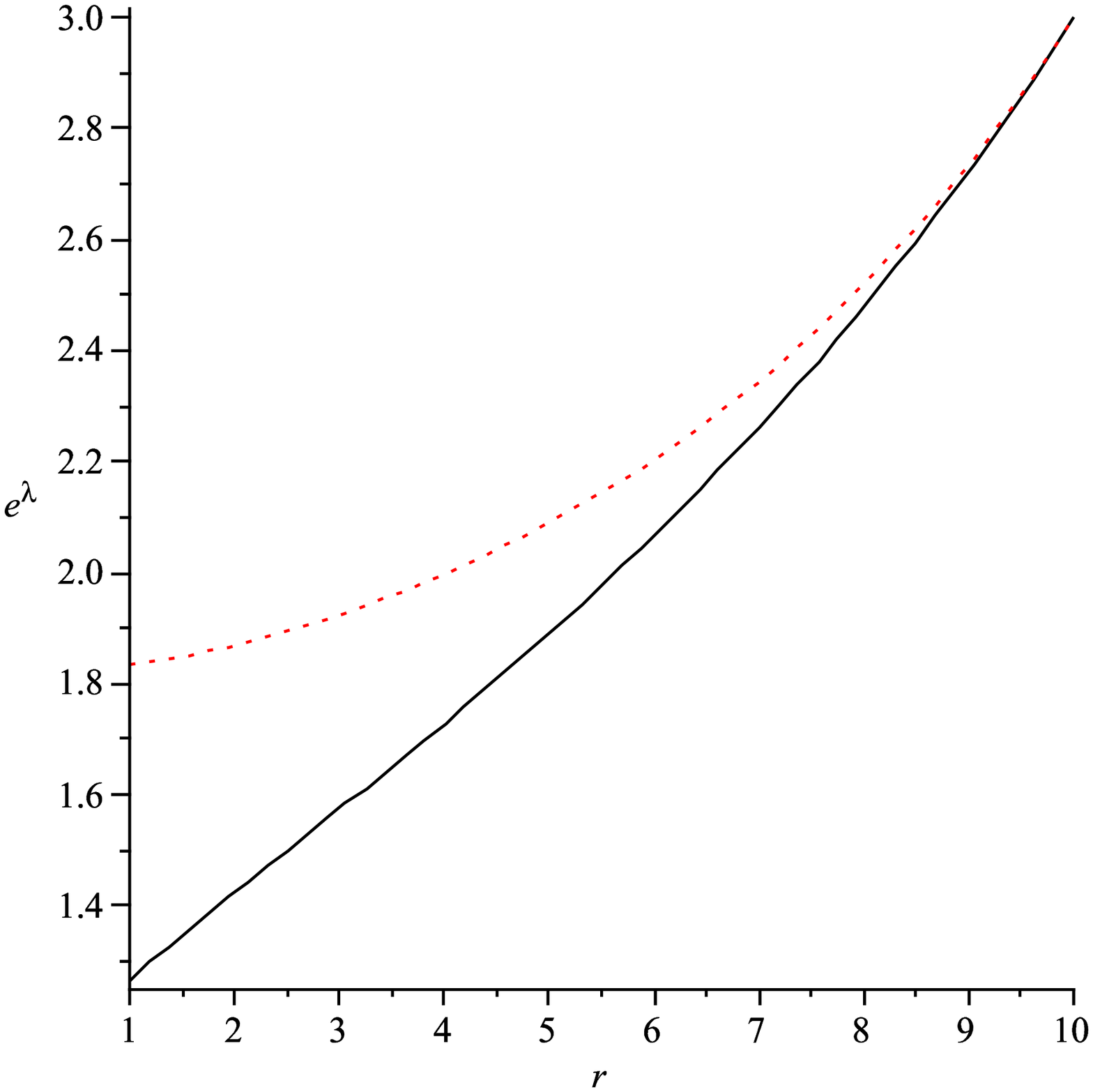}
    \caption{ Plot for the variation of $e^\lambda$  vs r( km ). Solid line and dotted line represent
    the case 1 and case 2 respectively for suitable choices of the parameters.
     }
    \label{}
\end{figure}
\begin{figure}
    \centering
        \includegraphics[scale=.33]{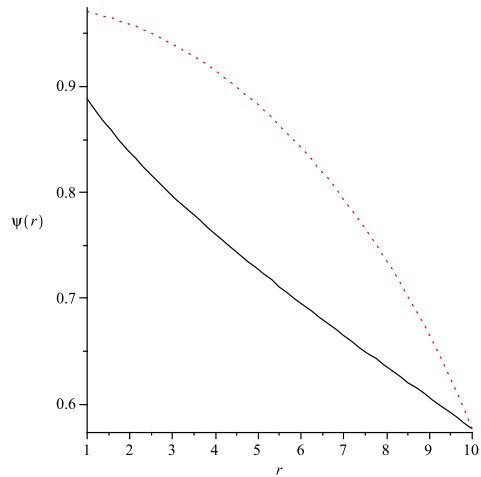}
    \caption{ Plot for the variation of conformal factor $\psi$  vs r( km ). Solid line and dotted line represent
    the case 1 and case 2 respectively for suitable choices of the parameters.}
    \label{}
\end{figure}
\begin{figure}
    \centering
        \includegraphics[scale=.33]{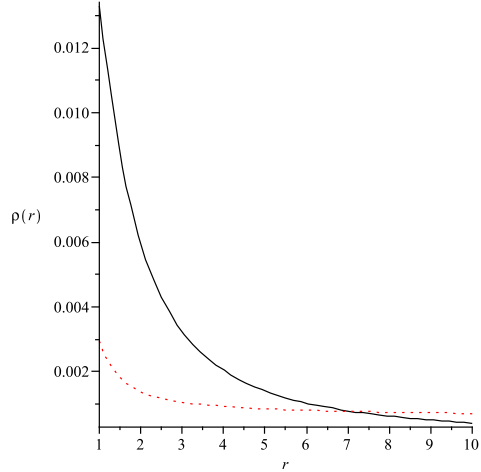}
    \caption{ Plot for the variation of energy density $\rho$  vs r( km ). Solid line and dotted line represent
    the case 1 and case 2 respectively for suitable choices of the parameters. }
    \label{}
\end{figure}
\begin{figure}
    \centering
        \includegraphics[scale=.33]{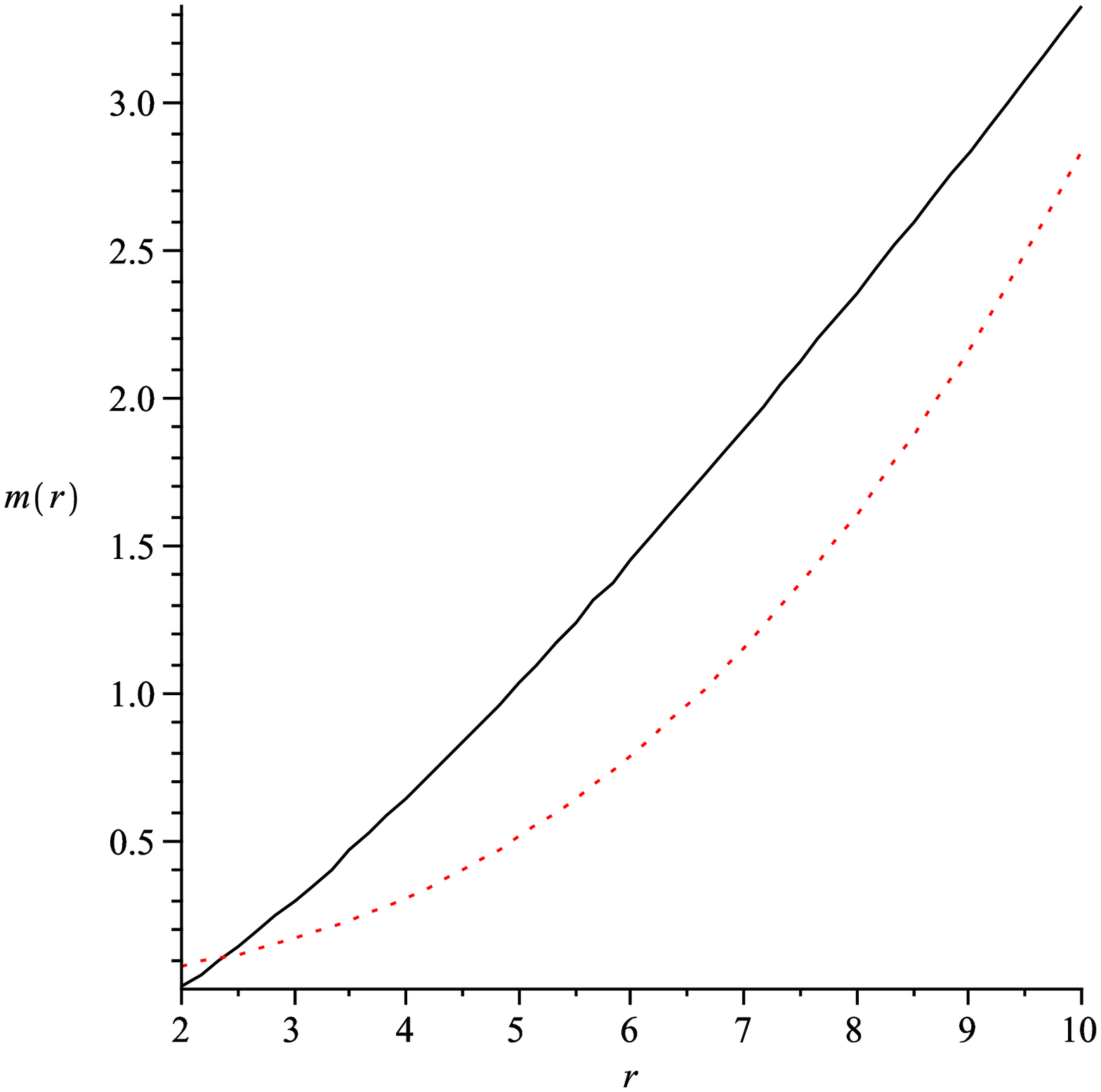}
    \caption{ Plot for the variation of mass $m(r)$  vs r( km ). Solid line and dotted line represent
    the case 1 and case 2 respectively for suitable choices of the parameters. }
    \label{}
\end{figure}
\begin{figure}
    \centering
        \includegraphics[scale=.33]{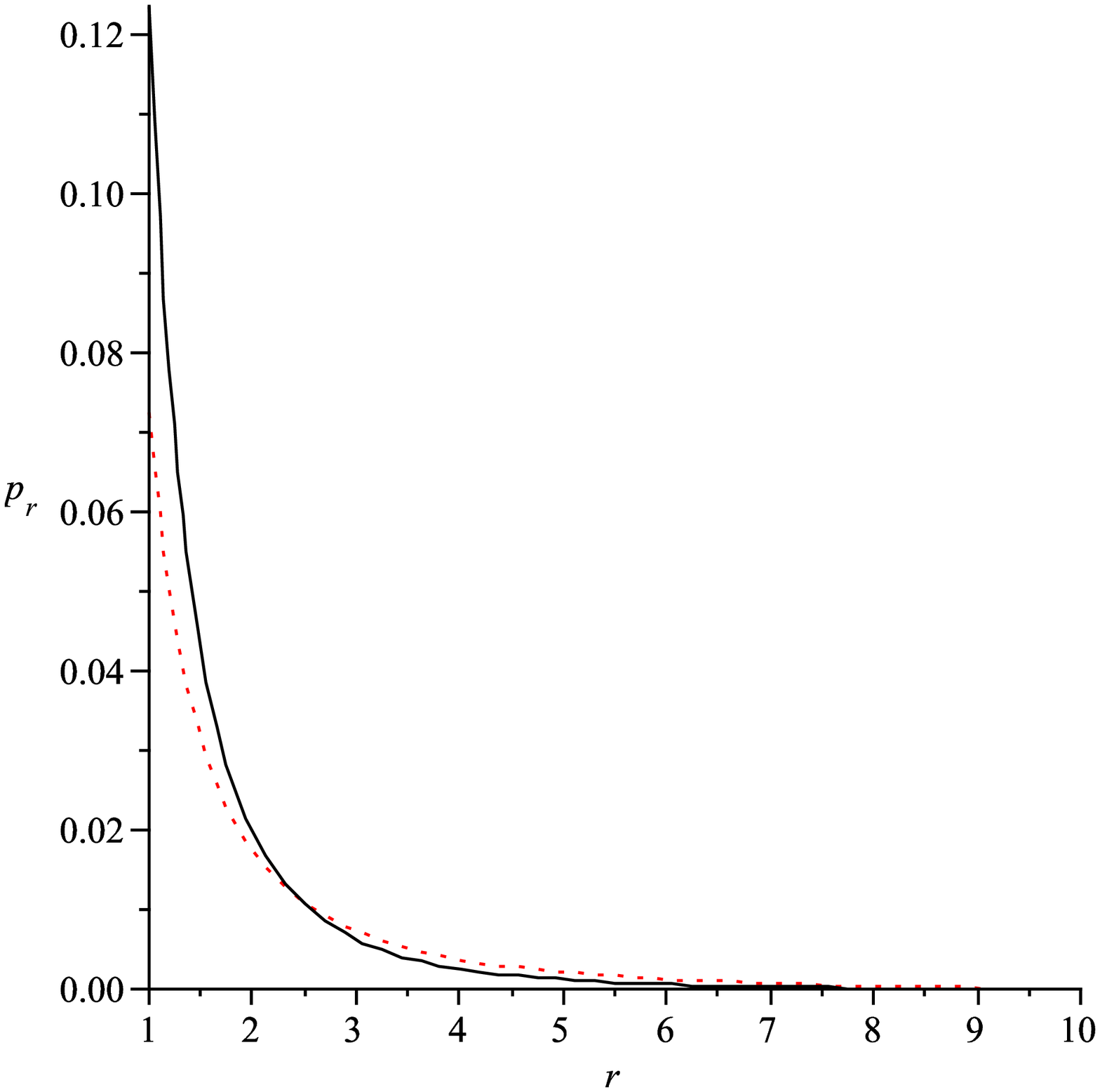}
    \caption{ Plot for the variation of the radial pressure $p_r$  vs r( km ). Solid line and dotted line represent
    the case 1 and case 2 respectively for suitable choices of the parameters. }
    \label{}
\end{figure}
\begin{figure}
    \centering
        \includegraphics[scale=.33]{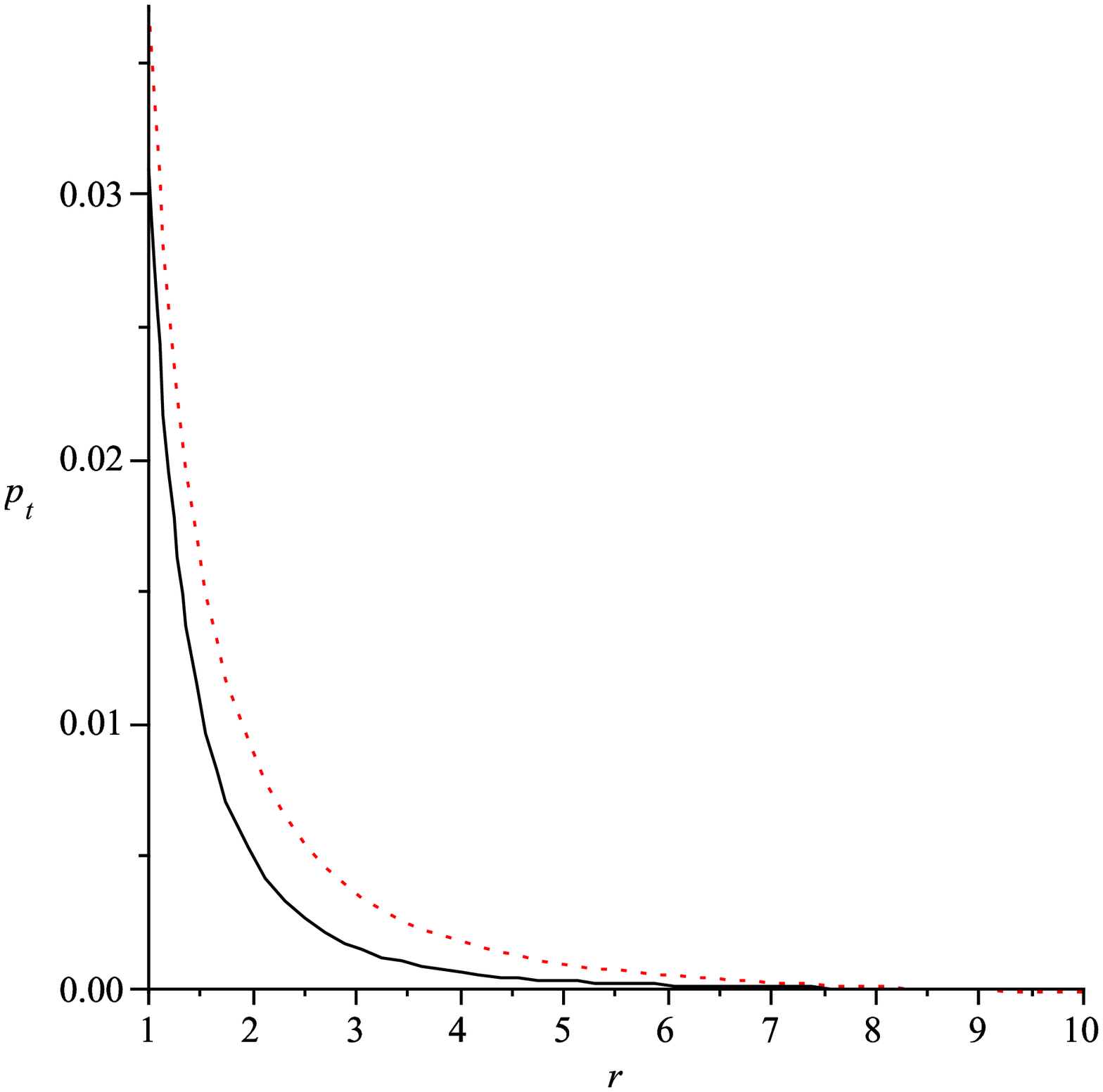}
    \caption{ Plot for the variation of the transverse pressure $p_t$    vs r(km). Solid line and dotted line represent
    the case 1 and case 2 respectively for suitable choices of the parameters. }
    \label{}
\end{figure}
\begin{figure}
    \centering
        \includegraphics[scale=.33]{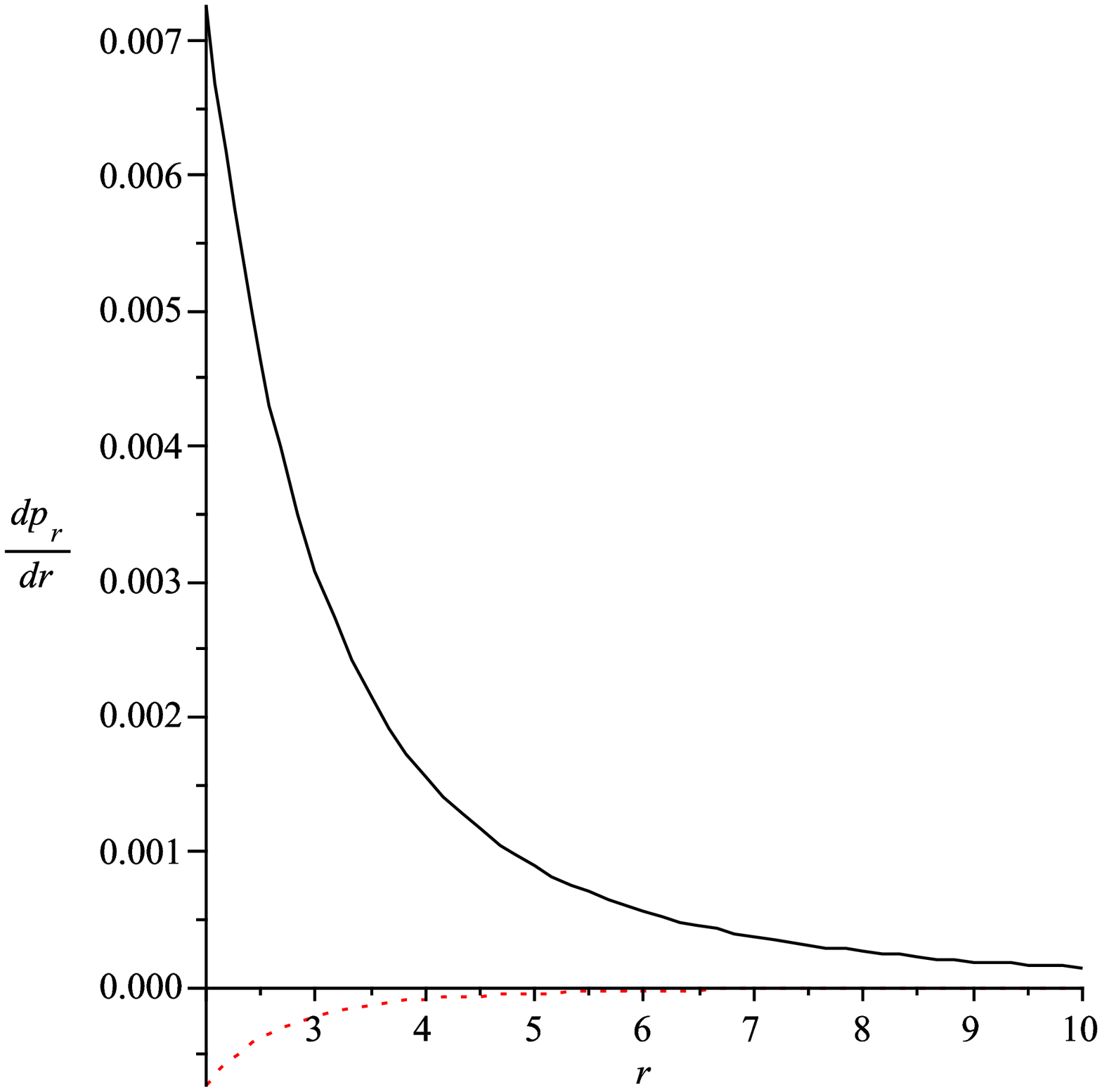}
    \caption{ Plot for the variation of the  gradient of radial pressure   $\frac{dp_r}{dr}$ vs r(km). Solid line and dotted line represent
    the case 1 and case 2 respectively for suitable choices of the parameters. }
    \label{}
\end{figure}
\begin{figure}
    \centering
        \includegraphics[scale=.33]{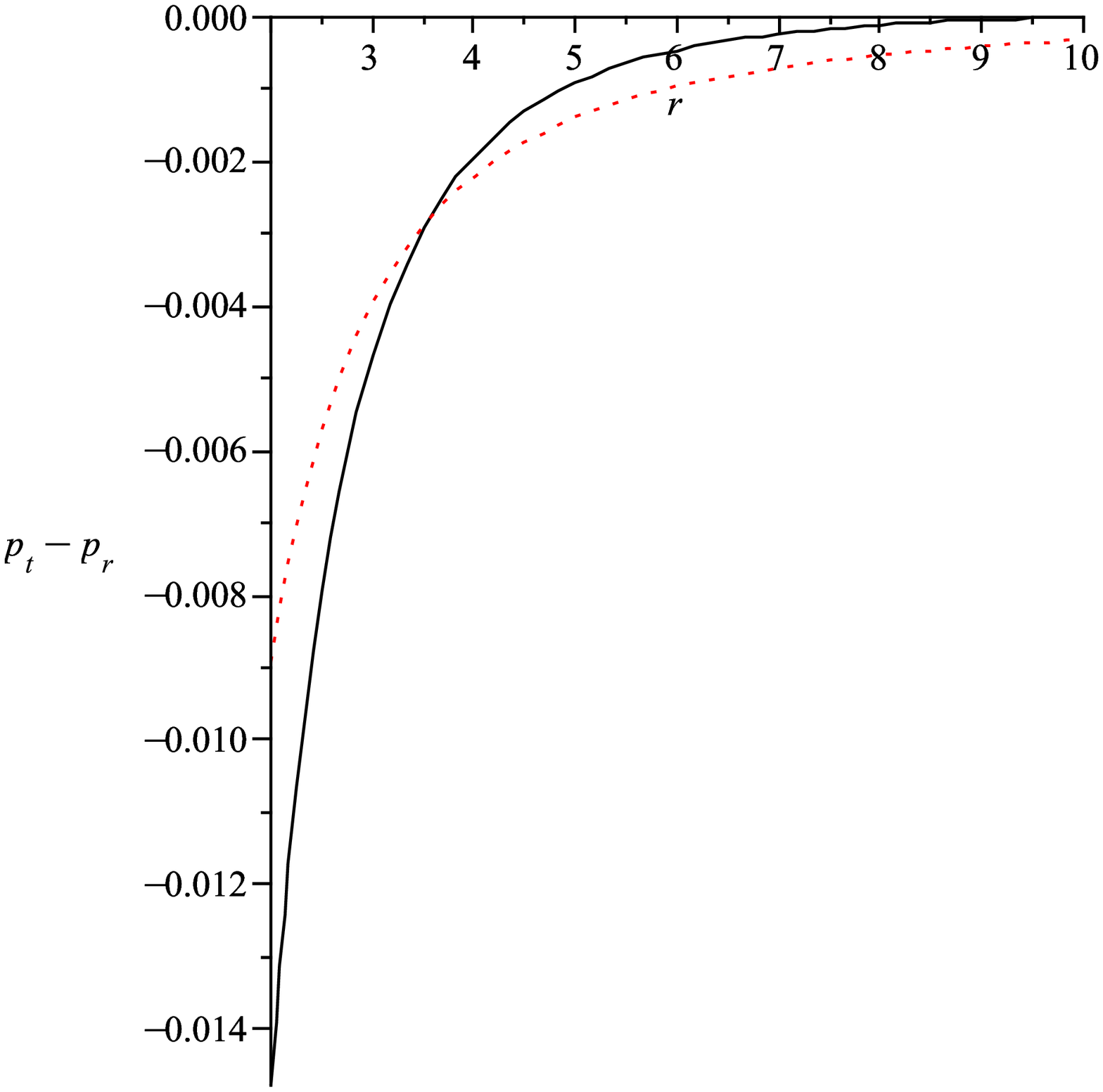}
    \caption{ Plot for the variation of the pressure anisotropy  $p_t - p_r$   vs r( km ).Solid line and dotted line represent
    the case 1 and case 2 respectively for suitable choices of the parameters.}
    \label{}
\end{figure}

\end{document}